# On Network Proximity in Web Applications

Dmitry Namiot[1], Manfred Sneps-Sneppe[2]
[1]*Lomonosov Moscow State University, Moscow, Russia*
[2]*Ventspils University College, Ventspils, Latvia*
[1]dnamiot@gmail.com; [2]manfreds.sneps@gmail.com

*Abstract*—In this paper, we discuss one approach for development and deployment of web sites (web pages) devoted to the description of objects (events) with a precisely delineated geographic scope. This article describes the usage of context-aware programming models for web development. In our paper, we propose mechanisms to create mobile web applications which content links to some predefined geographic area. The accuracy of such a binding allows us to distinguish individual areas within the same indoor space. Target areas for such development are applications for Smart Cities and retail.

## I. INTRODUCTION

Our paper deals with mobile web presentations of location-based services. How can we present some local (attached to a certain geographical location) information to mobile users? We are talking about programming (creating) mobile web sites, which content pages correspond to the current location of the mobile user. The traditional scheme is very straightforward. We have to determine the user's location and then create a dynamic web page, the issuance of which is clearly defined by specific geographical coordinates. For example, geo-location is a part of HTML5 standard [1].

As soon as web application obtained (as per user permission, of course) geo-coordinates, it can build a dynamic web page, which content depends on the current location (content is associated with obtained location). Technically, we can render our dynamic page on the client side (right in the browser), when application requests data from server via some asynchronous calls (AJAX) [2], or right on our server (in some CGI-script). In both cases obtained location info is used as a parameter either to AJAX script or to CGI script. For some of the applications (classes of applications), we may use several location-related datasets (e.g., so-called geo-fence [3]), but the common principles are similar. It is so called Location Based Services (LBS) [4].

There are different methods for obtaining location information for mobile users [5]. Not all of them use GPS (GLONAS) positioning actually. Alternative approaches use Wi-Fi, Cell ID, collaborative location, etc. [6]. The above-mentioned geo-location in HTML5 has been a wrapper (interface) for location service. For the most of LBS, their top-level architecture is standard. LBS use obtained location info as a key for any database (data store) with location-dependent data. Location info is actually no more than a key for linking physical space (location) and virtual (e.g., coupon for the store). Only a small number of services actually use the coordinates. The typical example is indoor location based services. The paradigm "Location first" requires a digital map for an indoor space. This map should be created prior to the deployment, and it should be supported in an actual state during service's life time. On the other hand, there is a direction, called context-aware computing. In context-aware computing (ubiquitous computing) services can use other information (not related to geographic coordinates) as the "characteristics" of a user's location. Simplistically, the context is any additional information on the geographical location [7-8]. In this case, additional information (context), with the presence of certain metrics can serve as a unique (up to a certain approximation, of course) feature of a user's location. Or, in other words, we can substitute geo-location with context identification. Why might it be necessary? The typical example is indoor LBS [9]. Traditional geo-positioning can be difficult and positioning accuracy may be insufficient to distinguish the position of the mobile subscriber within the same premises. And yet, it is the distinction between positions within the same space (buildings) may be important for all kinds of services (for example, the buyer is located on the first or second floor of the hall).

Actually, it is a starting point for new approaches in LBS architecture, when the stage with obtaining (detecting) location info could be completely eliminated. Indeed, if location info is no more than a key for some database, then why do not replace geo-keys (e.g., latitude and longitude) with context-related IDs? It is sufficient to identify context and use this identification to search data.

The rest of the paper is organized as follows. In section II, we describe context identification. In section III, we describe how this identification could be used in web programming. In section IV, we discuss the generic approaches for incorporating sensing information into web pages.

## II. NETWORK PROXIMITY

One of the widely used methods for the identification of context is the use of wireless network interfaces of mobile devices (Wi-Fi, Bluetooth). The reasons for this are straightforward. On the first hand, these interfaces are supported in all modern smart-phones. Secondly, for obvious reasons, monitoring of network interfaces is directly supported and executed by the mobile operating systems. Therefore, a survey of network interfaces on the application level can be simplified and not cause additional power consumption, as compared with, for example, a specially organized monitoring for the accelerometer.

Information received through the network interface is used to estimate the proximity of the mobile user to the elements of the network infrastructure (network proximity [10]). Note, that other mobile devices can act as these elements too (e.g., Wi-Fi access point, opened right onto mobile phone [11]).

The classical form for collecting data about Wi-Fi devices are so-called Wi-Fi fingerprints sets [12]. Wi-Fi fingerprints are digital objects that describe availability (visibility) for network nodes. Their primary usage is navigation related tasks. The alternative approach lets users directly associate some data chunks with existing (or artificially created) network nodes. In other words, it is a set of user generated links between network nodes and some content that could be used by those in proximity to networks nodes. This approach is presented in SpotEx project and associated tools [13-14]. SpotEx lets users create a set of rules (logical productions) for linking network elements and available content. A special mobile application (context-aware browser) is based on the external set of rules (productions, if-then operators). The conditional part of the each rule includes predicates with the following objects:

identity for Wi-Fi network (name, MAC-address)
RSSI (signal strength),
time of the day (optionally),

In other words, it is a set of operators like this:

*IF AccessPointIsVisible ('Café') THEN*
  *{ show content for Café }*

Block *{ show content for Café }* is some data (information) snippet presented in the rule. Each snippet has got a title (text) and some HTML content (it could be simply a link to any external site for example). Snippets could present coupons/discount info for malls, news data for campuses, etc. The context-aware browser (mobile application) maps current network environment against existing database, detects relevant rules (fires them) and builds a dynamic web page. This web page is presented to a mobile user in proximity.

In fact, even the name of the application (context-aware browser) suggests the movement of this functionality in a mobile browser. This would eliminate the separate rule base as well as the special (separate) application. In fact, the standard mobile browser should play a role of this application. Rules for the content (data snippets) must be specified directly on the mobile web pages. And data snippets itself are HTML code chunks anyway.

As applied implementations, we can mention, for example, Internet of Things applications [15-16]. The usage is very transparent. Data snippets (data, presented to mobile users) depends on visibility for some Wi-Fi access points. It lets us specify the positions for mobile users inside of some building (campus, etc.) Mobile users will see different information for different positions. And this approach does not use geo-coordinates at all.

The next interesting direction is EU project FI_WARE [17]. Integration with the FI-CONTENT platform is one of the nearest goals.

III. INFORMATION SERVICES

Technically, for the reuse of information about network proximity, we can talk about the two approaches.

At the first hand, the implementation of a mobile browser can follow the same ideology that supports geo-coding in HTML5 [18]. How does it work?

*<script>*
*function getLocation()*
  *{*
    *if (navigator.geolocation)*
    *{*
*//  interface function*
*navigator.geolocation.getCurrentPosition(showPosition); }*
  *}*

*// user-defined callback*
*function showPosition(position) {*
  *var latitude = position.coords.latitude;*
  *var longitude = position.coords.longitude;  }*
*</script>*

A function from browser's interface

  *navigator.geolocation.getCurrentPosition()*

accepts as a parameter some user-defined callback (another function). The callback should be called as soon as geo-location is completed. Obtained data should be passed as parameters. Note, that the whole process is asynchronous.

By the analogue with the above-mentioned model, a mobile browser can add a new interface function. E.g.,
  *getNetworks()*
this function will accept a user-defined callback for accumulating network information (current fingerprint). A good candidate for data model is JSON. The browser will pass fingerprint as a JSON array to a user-defined callback. Each element from this array describes one network and contains the following information:

  SSID -  name for access point
  MAC -  MAC-address
  RSSI -  signal strength

Note, that scanning networks is an asynchronous process in mobile OS. So, callback pattern is a good fit for this.

Firefox OS is closest in ideology to this approach [19]. Here is an example from the technical manual:

*interface WiFiManager {*
*// request.result set to JS array of wifi networks in range*
 *DOMRequest getNetworks();*

*// request fires success if*
*// successfully able to connect, error otherwise any*
*//connected; JSON object contains info on the connected*
*// network*
 *DOMRequest connectTemp(any parameters);*
 *int signalStrength;*

*// Fires event when we connect to a new WiFi network*

*Function onconnect;*

*// Fires event when we disconnect from a new*
*// WiFi network*
*Function ondisconnect;*

*// Fires event signal strength changes*
*Function onsignalstrengthchange;*
*}*

JSON object, returned by getNetworks() function, contains the following info: name (SSID), MAC-address, signal strength (RSSI) and security protocol.

Also, Firefox OS offers Bluetooth API [20]. It has got the similar ideology, but there is no general unifier (e.g., even fields for objects are different). It should be possible, of course, to create some unified wrapper (shell), which will give a general list of networks. But it is not the biggest problem. The biggest problem (we are not mentioning here the own prevalence and popularity for Firefox OS) is the status for both APIs. Wi-Fi API has just been scheduled yet. At the same time, the Bluetooth API exists, but it is declared preferred (privileged). Privileged APIs can be used by the operating system only. So, it could not be used in applications. The reason for this solution is security. API combines both network scanning and network connection (data exchange). It is the wrong design by our opinion. APIs functionality should be separated. The above-mentioned SpotEx approach is not about the connectivity. Mobile OS should use two separate APIs: one for scanning (networks poll) and one for connecting. Polling for networks does not require data exchange. So, scanning API is safe, and it should not be privileged. It is simple – we should have *WiFiManager* interface (as is, and it could be privileged), and *WiFiScan* with only one function getNetworks():

 <script>
    *function callback_function(json_data ) { … }*
    *WiFiScan.getNetworks(callback_function);*
 </script>

The callback function can loop over an array of existing networks IDs and show (hide) HTML div blocks with data related (associated) to the existing (visible) networks. Actually, it is a fundamental question. Traditionally, wireless networks on mobile phones are used as networks. But they are sensors too. The fact that some network node is reachable (visible) is a separate issue. And it could be used in mobile applications even without the ability to connect to that node. It is the main idea behind SpotEx, and it is the feature (option) we suggest to embed into mobile browsers.

How can we present our rules for network proximity? As per our suggestion, each data snipped should be presented as a separate div block in HTML code. E.g., the above-mentioned example looks so:

 *<div id="Café_rule">*
   *show content for Café*
 *</div>*

We can use CSS styles to hide/show this block. And this CSS visibility attribute depends on the visibility of Wi-Fi (Bluetooth) nodes. Of course, CSS visibility could be changes in JavaScript. So, our rules could be implemented in JavaScript code. We can directly code predicates in our code, or describe their parts in CSS too. E.g.:

 *<div id="Café_rule" cond="Café_AP1 Café_AP2" >*
   *show content for Café*
 *</div>*

In this example, an additional attribute *cond* contains a list of Wi-Fi access point that should be visible for showing that block. HTML5 custom attributes are good candidates for new attributes [21]. It means also, that adding some set of rules to existing web page is no more that adding (including) some JavaScript code (JavaScript file).

In general, this approach could change the paradigm of designing mobile web sites. It eliminates the demand to make separate versions for local sites or events. It is enough to have one common site with local offers (events, etc.) placed in hidden blocks. Blocks will be visible to mobile users in a proximity of some network nodes. Local blocks visibility depends on the network nodes visibility and so, it depends on the current location of mobile users. E.g., for the above-mentioned example, mobile users opened Café site being physically present in the proximity of Café, will see different (additional) data compared with any regular mobile visitor.

Of course, single data source (just one web site) support simplifies (makes it cheaper) the maintenance during life time.

Web Intents [22] present the next interesting model for this approach. The Web Intents formation is a client framework (everything is executed in the browser) for the monitoring (polling) and building services interaction within the application. Interactions include data exchange and transfer of control. Some service (named code snippet) announces its readiness to support some operations. For example, a service can declare its ability to edit some text (images), send messages, etc. Application (custom code) requests a service for some action (edit, send, etc.). The executive system picks service based on its announcement. In our particular case, we are talking about Intent service that polls the network environment. Note, Web Intents are asynchronous (there is a standard callback function *onActivity*).

Web Intents form the core architecture of Android OS [23], but their future status is still unknown after some initial experiments from Google. We should note in this context a similar (by its concept) initiative from Mozilla Labs - Web Activities [24]. But the further status of this initiative is also unclear.

The next possible toolbox is seriously underrated in our opinion. It is a local web server. The first implementation, as far as we know, refers to the Nokia [25]. In the original paper, authors port Nokia Apache server on the S60 platform. In our opinion, this is one of the most promising areas for communicating with phone sensors. Most of the services (applications) need read-only access to data from sensors. It is a data programming interface (DPI), rather than an application programming interface (API). So, this local server can simply provide a set of CGI scripts for reading data from sensors. Each script can return data in JSONP format. So, any polling for sensors (including network proximity) looks like a simple JavaScript code inclusion:

*<script type="text/javascript" src="http://localhost:8080/getNetworks?callback=f" ></script>*

Here *f* is the name of the function that is called at the completion of the sensors polling. This function will accept the above-mentioned JSON array with information about the network nodes. As you see, this approach is clearly exploiting the fact, that in practice, for most purposes, access to sensors is read-only. In our case, network proximity is a perfect example of Data Program Interface as the most of the other tasks in context-sensitive programming. It means, by the way, that sometimes attempts to create a comprehensive API, like the above-mentioned Web Bluetooth of Firefox OS, actually slows down the process. They can create non-existing problems (e.g., security concerns for read-only access) and take great care about not used features.

Note, that for the mobile users where the local web server is not available (or sensors are not available) callback's execution will be simply omitted. In other words, we can support one source (one web page) for all users again.

The next possible idea resembles in some ways the old projects with WAP (Wireless Access Protocol) – Figure 1.

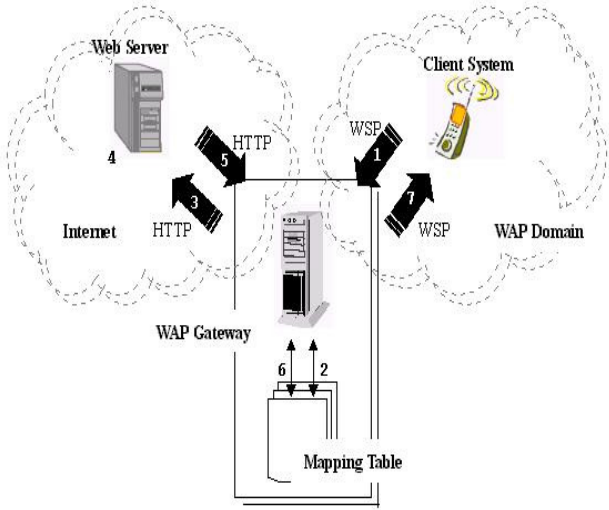

Figure1. WAP proxy [26]

In this case, a mobile device used some intermediate server (WAP Gateway) for access to internet resources. This intermediate server should be able to collect sensing information (including network sensors). Internet service will get sensing info from our proxy. In other words, any access from a mobile web browser to the internet should be passed via proxy. And the obvious candidate for the proxy's location is the mobile device itself. E.g., the above mentioned on-board web server from Nokia (actually, it was the modified Apache server) can play a role of the proxy. This embedded server (as a mobile application) can access to the local sensors and, at the same time, intercepts outgoing HTTP requests, enrich them with obtained (saved during a session) sensing information and pass HTTP request to the target site.

## IV. A GENERIC APPROACH FOR WEB SENSING

In this part, we would like to discuss the more generic approach (approaches) for embedding sensing information into web pages.

In this connection, we should mention the following sources. On the first hand, it is W3C Semantic Sensor Network Incubator Group (SSN-XG) [27]. Firstly, this group is developing an ontology to describe sensors and their device, system and platform related attributes. At the second, it develops a semantic markup and recommends methods to use ontology to describe the data available based on the existing models such as the Open Geospatial Consortium's (OGC) Sensor Web Enablement (SWE) [28] standards. SWE standards that have adopted by the OGC membership include the following elements:

1. Observations & Measurements Schema (O&M). It is the XML Schema for encoding observations and measurements from a sensor, both archived and real-time.

2. Sensor Model Language (SensorML). It is a set of standard models and XML Schema for describing sensors, systems and processes; provides information needed for discovery of sensors, location of sensor observations, processing of low-level sensor observations, and listing of their properties. Bluetooth node could be described as a "standard" sensor with observable properties "MAC-address" and "RSSI". In the terms of Sensor Ontology [29], it should be a simple element.

3. Transducer Markup Language (TransducerML or TML). It is the conceptual model and XML Schema for describing transducers and supporting real-time streaming of data to and from sensor systems.

4. The Sensor Observations Service (SOS) provides a standard web service interface for requesting, filtering, and retrieving observations and sensor system information. This is the proxy between a client and an observation repository or near real-time sensor channel. Actually, it is like a standard interface for the above mentioned "on-board" web proxy.

5. The Sensor Planning Service (SPS) is similar to SOS, but used for the management. It provides a web service interface for requesting user-driven acquisitions and observations. This is the broker between a client and a sensor collection management environment.

6. The Sensor Alert Service (SAS) – provides a standard web service interface for publishing and subscribing to alerts from sensors.

7. The Web Notification Services (WNS) – provides a standard web service interface for asynchronous delivery of messages or alerts from SAS and SPS web services.

The whole scheme looks over-engineered for a network proximity task. We think that HTML5 micro-data approach is much more promising here (for this particular task, of course).

The next approach we should mention in this context is W3C Web Applications Working Group [30]. This Group is working on creating specifications that enable improved client-side application development on the Web. This development includes specifications both for APIs for client-side development and for markup vocabularies for describing and controlling client-side application behavior.

This development is a part of the Rich Web Clients Activity in the W3C Interaction Domain [31]. In particular, W3C Web Applications working group targets An Application Programming Interface (API) in the forms of client-side script APIs, for use in browsers and similar user agents (as opposed to server-side APIs, for example).

The component model for the Web proposed by this group includes the following elements.

1. Templates, which define chunks of the markup that are inert, but can be activated for use later. Actually, the above-mentioned div-blocks are perfect examples of templates.

2. Decorators, which apply templates, based on CSS selectors to affect rich visual and behavioral changes to documents.

3. Custom Elements, which let authors (developers) define their own elements, with new tag names and new script interfaces. There are so-called widgets.

Widgets are defined as full-fledged client-side applications that are authored using technologies such as HTML, then packaged for distribution and, typically, downloaded and installed on a client machine or device where they run not only as stand-alone applications, but also embedded into Web pages and run in a Web browser [32].

Of course, it is just an interface. This group develops the way browsers will communicate with external data sources. For the actual data gathering support, we have to investigate the development of W3C Ubiquitous Web Domain Group [33]. This Group is focusing on technologies to enable Web access for anyone, anywhere, anytime, using any device. This includes Web access from mobile phones as well as other emerging environments such as consumer electronics, interactive television, and even automobiles [33]. For example, The Device APIs and Policy Working Group are creating client side programming interfaces to enable Web applications and widgets to access device services, including the calendar, contacts, camera, etc. The group will also provide a framework for expressing security policies to govern access to these APIs. One of the examples, we could be interesting in is *Proximity Events* interface.

The *DeviceProximityEvent* interface provides web developers information about the distance between the hosting device and a nearby object. The *UserProximityEvent* interface provides web developers a user-agent- and platform-specific approximation that the hosting device has sensed a nearby object.

This is achieved by interrogating a proximity sensor of a device, which is a sensor that can detect the presence of a physical object without physical contact. Not all devices contain a proximity sensor, and when there is no proximity sensor, this API is still exposed to the scripting environment, but it does nothing [34]. This approach supposed to support so-called proximity sensors. They can use radiation (e.g., an infrared light or a magnetic field), certain material properties can interfere with the sensor's ability to sense the presence of a physical object. The spec is directly mentioned that objects that can interfere with a sensor include, but are not limited to, the material's translucency, color, temperature, chemical composition, and even the angle at which the object is reflecting the radiation back at the sensor. As such, proximity sensors should not be relied on as a means to measure distance: the only thing that can be deduced from a proximity sensor is that an object is somewhere in the distance between the minimum sensing distance and the maximum sensing distance with some degree of certainty. Actually, this definition covers the network proximity too. But as far as we know, at this moment nobody threats wireless network nodes as potential things in proximity tasks.

There is a sub-division of this group which targets Network API, but their documents declare the interest on the network connectivity only. At the same time, as we explained above, network proximity does not require the connectivity at all.

In fact, in this paper, we would like to draw attention to the fact that the network proximity deserves a separate API. For example, there is an existing initiative for NFC related Web API, and we see no reasons why more widespread Bluetooth and Wi-Fi have not such development.

As a workaround and prototype for this development, we can present a custom *WebView* for Android. On Android platform is possible to access from JavaScript to Java code for a web page, loaded into *WebView* control. Java code will provide a list of nearby network nodes (calculate the network fingerprint). The key moment here is the need for an asynchronous call from JavaScript, because scanning for wireless networks in Java is the asynchronous process. Let us describe this approach a bit more detailed.

On Android side we activate JavaScript interface:

```
public void onCreate(Bundle savedInstanceState) {
super.onCreate(savedInstanceState);
    WebView webView = new WebView(this);
    setContentView(webView);
    WebSettings settings = webView.getSettings();
settings.setJavaScriptEnabled(true);
    webView.addJavascriptInterface(new MyJavascriptInterface(), "Network"); }
```

Now we can describe our Java code for getting network fingerprint. As a parameter, we will pass a name for callback function in JavaScript.

```
@JavascriptInterface
public void getNetworks(final String callbackFunction) {  }
```

We skip the code for network scanning and demonstrate the final part only. As soon as a fingerprint in obtained, we can present it as JSON array and invoke our callback:

```
webView.loadUrl("javascript:" + callbackFunction + "('" + data + "')");
```

And on our web page, we can describe our callback function and call Java code:

```
function f_callback(json) {  }
Network.getNetworks("f_callback");
```

This approach lets us proceed network proximity right in JavaScript (in other words, right on the web page). Actually, by the similar manner we can work with other sensors too. It is so-called Data Program Interface [35]. We would like to see something similar as a standard feature in the upcoming versions of Android.

## V. CONCLUSION

The paper discusses the use of information about the network environment to create dynamic web pages. We propose the several approaches to the implementation of a mobile browser that can handle data on a network (network proximity) to provide users with information tied to the current context. Also, we considered possible implementation details. The basic idea is to separate the functional for scanning network information and real data exchange.